\documentclass{emulateapj}

\usepackage{natbib} 
\begin{document}

\title{The DEEP2 Galaxy Redshift Survey: The Evolution of Void Statistics 
from $z\sim1$ to $z\sim0$}

\author{
Charlie Conroy\altaffilmark{1},
Alison L. Coil\altaffilmark{1},
Martin White\altaffilmark{1,3},
Jeffrey A. Newman\altaffilmark{2}, 
Renbin Yan\altaffilmark{1},
Michael C. Cooper\altaffilmark{1},
Brian F. Gerke\altaffilmark{3},
Marc Davis\altaffilmark{1,3}, 
David C. Koo\altaffilmark{4}
}

\altaffiltext{1}{Department of Astronomy, University of California,
Berkeley, CA 94720 -- 3411} 
\altaffiltext{2}{Hubble Fellow, Lawrence Berkeley National Laboratory, 
1 Cyclotron Road, Berkeley, CA 94720}
\altaffiltext{3}{Department of Physics, University of California,
Berkeley, CA 94720 -- 3411}
\altaffiltext{4}{University of California Observatories/Lick
Observatory, Department of Astronomy and Astrophysics, University of
California, Santa Cruz, CA 95064}

\begin{abstract}
We present measurements of the void probability function (VPF) at
$z\sim1$ using data from the DEEP2 Redshift Survey and its evolution
to $z\sim0$ using data from the Sloan Digital Sky Survey (SDSS).  We measure
the VPF as a function of galaxy color and luminosity in both surveys
and find that it mimics trends displayed in the two-point correlation
function, $\xi$; namely that samples of brighter, red galaxies have
larger voids (i.e. are more strongly clustered) than fainter, blue
galaxies.  We also clearly detect evolution in the VPF with cosmic
time, with voids being larger in comoving units at $z\sim0$.   We find
that the reduced VPF matches the predictions of a `negative binomial'
model for galaxies of all colors, luminosities, and redshifts studied.
This model lacks a physical motivation, but produces a simple analytic
prediction for sources of any number density and integrated two-point
correlation function, $\bar{\xi}$.  This implies that differences in
the VPF  across different galaxy populations are consistent with being
due entirely  to differences in the population number density and
$\bar{\xi}$.  We compare the VPF at $z\sim1$ to $N$-body $\Lambda$CDM
simulations and find good agreement between the DEEP2 data and mock
galaxy catalogs.  Interestingly, we find that the dark matter {\it
particle} reduced VPF follows the physically motivated `thermodynamic'
model, while the dark matter {\it halo} reduced VPF more closely
follows the negative binomial model. The robust result that all galaxy
populations follow the negative binomial model appears to be due to
primarily to the clustering of dark matter halos.  The reduced VPF is
insensitive to changes in the parameters of the halo occupation
distribution, in the sense that halo models with the same $\bar{\xi}$
will produce the same VPF.  For the wide  range of galaxies studied,
the VPF therefore does not appear to provide useful constraints on
galaxy evolution models that cannot be gleaned from studies of
$\bar{\xi}$ alone.

\end{abstract}

\keywords{galaxies: evolution --- galaxies: dark matter --- galaxies: clustering}

\section{Introduction}
Voids are some of the most striking large-scale features of the
Universe.  Historically, their study can be loosely grouped into two
categories: finding individual voids or using a statistical approach
(see \citet{Rood88} for a detailed review of the history of void
studies).  The first focuses on identifying individual voids with
sophisticated void-finding algorithms \citep{Kauffmann91, Kauffmann92,
Elad96, Ryden96, ElAd97b, Aikio98, Sheth03, Patiri05} that allow voids to have
any convex shape.  The properties of galaxies in voids can then be
studied, including their color distribution, luminosity function,
concentrations and star-formation rates
\citep{Grogin99,Grogin00,Rojas04, Hoyle05, Rojas05}.  \citet{Hoyle04}
analyze the recently-completed 2dF Galaxy Redshift Survey and find
that void galaxies are, on average, bluer and show evidence for more
recent star-formation than the full 2dF galaxy population.
Properties of voids themselves can also be studied, including
characteristic sizes, mean ellipticities, and radial density profiles.
Many of these observational quantities, such as size measurements, can
only be interpreted with carefully constructed mock galaxy catalogs.
Extensive theoretical work has been carried out  concerning void size
distributions and density profiles \citep{Sheth04,  Benson03,
Colberg05, Shandarin04, Patiri04}, void halo properties
\citep{Antonuccio-Delogu02,  Gottlober03, Goldberg04, Patiri04, Colberg05},
formation of galaxies  within voids \citep{Mathis02}, and the
properties of galaxies in voids \citep{Benson03}.  The question of
whether simulations can produce voids as large and as empty as  voids
seen in the observed Universe remains unanswered, and can potentially
provide powerful  constraints on galaxy formation and cosmology
\citep{Peebles01}.

The second approach to voids is somewhat more statistical and is in
many ways complementary to the first.  Studies of this type focus
primarily on the void probability function (VPF) which is defined as
the probability that a sphere of a given size centered on a random
point in the  survey volume contains no galaxies.  Statistical void
distributions in principle offer a wealth of information, as they can
be related to the entire hierarchy of galaxy correlation functions
\citep{White79}.  Yet, due to the difficulties in interpreting void
statistics, they have not received as much attention as more
conventional statistical measures of large-scale structure such as the
two-point correlation function.  Despite these difficulties, the
theoretical framework underlying void statistics has been developed in
detail \citep{White79, Fry84a, Fry85, Fry86, Otto86, Fry88, Sheth96,
Balian89}, and has been extensively studied in simulations
\citep[e.g.][]{Fry89, Ghigna94, Ghigna96, Ryden96, Kauffmann97,
Schmidt01, Berlind02,  Benson03}.

Observationally, voids statistics have been investigated in almost
every major galaxy redshift survey, including the CfA
\citep{Maruogordato87, Vogeley91,Vogeley94, Mo90}, SSRS
\citep{Gaztanaga93}, PSCz \citep{Hoyle02}, IRAS \citep{Elad97a}, LCRS
\citep{Muller00} and most recently, the 2dF survey \citep{Hoyle04,
Croton04, Patiri05}.  Results from the latter survey are
representative of an emerging consensus; subsamples of brighter and/or
redder galaxies contain larger voids than samples of fainter and/or
bluer galaxies.  This is interpreted as brighter and/or redder
populations being more strongly clustered than fainter and/or bluer
subsamples.  These trends are also reflected in the two-point
correlation function of galaxies, $\xi$, both at low
\citep[e.g.,][]{Zehavi02} and moderate \citep[e.g.,][]{Coil04a}
redshifts.   We note that some studies of voids have not taken into
account the fact that galaxy populations have different number
densities which will strongly affect void statistics; hence
differences in the VPF across galaxy populations might as easily be
attributable to the luminosity function as to varying clustering
strengths.  As brighter galaxies are rarer than fainter galaxies, even
in volume-limited samples, brighter  samples will have higher void
probabilities.   A critical question we wish to address here is the
extent to which void statistics are governed by low-order clustering
statistics.  It is therefore important to separate the effects of
clustering from the effects of number density on the VPF.

Void statistics can also be used to probe the relation between
galaxies and dark matter.  There are strong theoretical arguments that
show that the clustering strength of dark matter and galaxies should
be different
\citep[e.g.,][]{Kaiser84,Bardeen86,Efstathiou88,Cole89,Mo96}.
Galaxies  initially form in high density peaks of the dark matter
distribution and hence are ``biased'' tracers of the dark matter at
high redshift.  Baryonic physics  and cosmic evolution also lead one
to expect that the biasing between  galaxies and dark matter should,
in principle, be a function of scale, redshift, and galaxy properties
such as color and luminosity.  Many of these expectations have been
borne out in observations \citep[see for
example][]{Davis76,Loveday95,Zehavi02, Madgwick03c,Coil04a}.

By studying voids in observational samples and comparing to dark
 matter simulations, one can hope to probe the galaxy bias in a unique
 way.  The VPF has previously been studied in simulations where a
 biasing prescription was assumed in order to place galaxies in the
 dark matter distribution \citep[e.g.,][]{Little94, Kauffmann97,
 Berlind02,  Ghigna94,Ghigna96}.  \citet{Kauffmann97} find that the
 VPF has limited utility in probing the bias, as the relation between
 dark matter and galaxies changes as a function of the sampling
 density of a galaxy survey.  \citet{Weinberg92} and \citet{ Little94}
 however, find that the VPF can be a powerful discriminator between
 various biasing schemes, but that it is relatively insensitive to the
 value of the linear bias factor $b\equiv(\xi_{gal}/\xi_{dm})^{1/2}$.
 For example,  the VPF is very different, for fixed $b$, when one
 statistically  identifies galaxies with overdensities in the initial
 density field  (`peaks biasing') versus identifying galaxies with
 overdensities in the final matter distribution (`density biasing').

More recently, the halo model
\citep[e.g.,][]{Seljak00,Peacock00,Cooray02,Kravtsov04} has emerged as
a useful prescription for placing galaxies within dark matter
simulations.  Instead of using a biasing scheme such as `peaks
biasing'  or `density biasing' to relate galaxies to the dark matter
distribution,  the halo model places galaxies within virialized dark
matter halos as a function of the halo mass.  Although in principle
the parameters of the halo model can evolve with cosmic time, there is
evidence to suggest that the model does not strongly evolve from
$z\sim1$ to $z\sim0$ \citep{Yan03}.  \citet{Berlind02} find that the
VPF can be  used to determine halo model parameters such as the
minimum dark matter halo mass, $M_{min}$, in which a galaxy of a given
luminosity may exist, and can hence provide useful new constraints on
the relation between galaxies and dark matter.  In this paper we
explore in detail the possibility of using the VPF to constrain the
halo model.

Until now, there has not been sufficient data to study the statistics
of voids at intermediate redshift ($z>0.3$).  From correlation
function measurements using the DEEP2 dataset at $z\sim1$
\citep{Coil04a} we know that galaxies were in general less strongly
clustered in the past; one might expect this trend to be reflected in
void statistics as well.  As more data becomes available at higher
redshifts, the importance of self-consistently investigating the
evolution of galaxy properties and statistics cannot be
overemphasized.  This entails using similar selection and analysis
techniques at different redshifts.  For this reason, we have chosen to
include in this study an analysis of the SDSS, which now has data for
$\sim200,000$ galaxies at $z<0.2$, to allow for robust conclusions
concerning the evolution of void statistics.

It should be kept in mind that the two approaches to the study of
voids outlined above use the term ``void'' in very different ways.
While the first approach identifies voids as large underdensities in
the galaxy distribution, allowing a void to assume any convex shape
and allowing galaxies to exist inside voids, the second approach
identifies voids as spherical regions in which surveyed galaxies are
totally absent.  Furthermore, while the first approach only locates
``unique''  voids, insofar as it does not allow smaller ``sub-voids''
to be  contained within larger voids, the second approach allows for
``voids''  to overlap (see $\S$\ref{s:deep} for a visualization of
this idea).  The statistical approach is only interested in the
question: what is the probability that a given volume element in the
universe is empty?  In this paper we are concerned with the second,
statistical approach, and hence we define voids in the latter sense.

The rest of this paper is organized as follows.  In $\S$\ref{s:thy} we
outline the theory behind void statistics and point out several
problems that arise with the theoretical underpinnings of voids.
$\S$\ref{s:meth} describes our methodology.  We statistically analyze
voids in mock galaxy catalogs built from N-body simulations in
$\S$\ref{s:mocks}; in $\S$\ref{s:deep} we present our results from the
DEEP2 galaxy survey at $z\sim1$ and in $\S$\ref{s:sdss} we analyze the
SDSS galaxy survey at $z\sim0$.  $\S$\ref{s:disc} discusses the
implications and relevance of our results.  Throughout this paper we
assume a flat concordance $\Lambda$CDM cosmology with $\Omega_m =
0.3$, $\Omega_{\Lambda} = 1-\Omega_m = 0.7$ and $H_0 = 100$ $h$ km
s$^{-1}$ Mpc$^{-1}$.

\section{Theoretical Background}\label{s:thy}

\subsection{General Considerations}
The void probability function (VPF) is defined as the probability of
finding no galaxies inside a sphere of radius $R$, randomly placed
within a sample.  The most common theoretical interpretation of the
VPF is as an infinite sum of the hierarchy of correlation functions of
the galaxy distribution \citep{White79,Sheth96}.  Specifically, for
spherical volume elements one may write the VPF as:
\begin{equation}
\label{eqn:vpf}
P_0(R) =
exp\left[\sum_{p=1}^\infty\frac{-\bar{N}(R)^p}{p!}\bar{\xi}_p(R)\right],
\end{equation}
where $R$ is the sphere radius, $\bar{N}$ is the average number of
galaxies within the sphere, and $\bar{\xi}_p$ is the volume averaged
$p$-point correlation function, where the volume average is defined by
\begin{equation}
\bar{\xi}_p\equiv\frac{\int\xi_pdV}{\int dV}.
\end{equation}

As $P_0$ depends on a recurring factor of $\bar{N}$, any meaningful
comparison of $P_0$ between populations requires a careful handling of
their number densities.  For many comparisons made in this study, we
choose to remove the dependency on $\bar{N}$ by randomly diluting the
samples to have similar number densities.  In this way we can isolate
the effects of clustering on the VPF.

The VPF takes a much simpler form if one uses the hierarchical
\emph{Ansatz},
\begin{equation}
\label{eqn:ansatz}
\bar{\xi}_p = S_p\bar{\xi}^{\,p-1}, \,\,\,\,\,\,\,\, p\geq3
\end{equation}
to relate the hierarchy of correlation functions to the two-point
function, $\bar{\xi}$. This allows for a complete and relatively
simple  description of the entire cosmological density field.  The
\emph{Ansatz} has been formally derived only for an $\Omega_m=1$
universe with the additional assumptions of stable clustering and
self-similarity \citep{Bernardeau02}, which are both known to be
invalid.  In the linear regime ($R\gtrsim15$ Mpc) perturbation theory
seems to validate the \emph{Ansatz}, although this paper is largely
confined to smaller scales.  So far,  the \emph{Ansatz} has not been
ruled out by observations \citep[see e.g.][]{Gaztanaga92,
Gaztanaga95,Baugh04}, although it is worth mentioning that Baugh et
al. had to remove the largest structure in their sample in order to
recover scale invariant $S_p$ values.  There is no  reason to believe
that the \emph{Ansatz} should hold in the quasi-linear to  strongly
nonlinear regimes.  Although historically the $S_p$ values were
assumed  to be scale invariant, nothing in the formalism developed
below requires  them to be.  With the hierarchical \emph{Ansatz} the
VPF becomes:
\begin{equation}
\label{eqn:fry}
P_0 =
exp\left[\sum_{p=1}^\infty\frac{-\bar{N}^p}{p!}S_p\bar{\xi}^{\,p-1}\right].
\end{equation}
\citet{Fry86} noted that Eqn.~\ref{eqn:fry} could be manipulated to
isolate the effects of the scaling coefficients ($S_p$).  Fry defined
the reduced void probability function, $\chi$:
\begin{equation}
\label{eqn:rvpf}
\chi \equiv -ln(P_0)/\bar{N}
\end{equation}
which, with the substitution of Eqn.~\ref{eqn:fry} becomes:
\begin{equation}
\label{eqn:chi}
\chi(\bar{N}\bar{\xi}) =
\sum_{p=1}^\infty\frac{S_p}{p!}(-\bar{N}\bar{\xi})^{p-1}.
\end{equation}
With this definition, $\bar{N}\bar{\xi}$ becomes the independent
variable, and hence only populations of galaxies with different $S_p$
values will have different reduced VPFs.  Although $\bar{N}\bar{\xi}$
will in general have a different dependence on $R$ for each galaxy
population, this quantity will always be an increasing function of $R$
since $\bar{N}\propto R^3$ and $\bar{\xi}\propto R^{-\gamma}$ with
$\gamma<3$ for all of the populations considered here.  We now briefly
summarize several models which  predict values for the $S_p$ values.

\subsection{Hierarchical Models}

\begin{figure}
\plotone{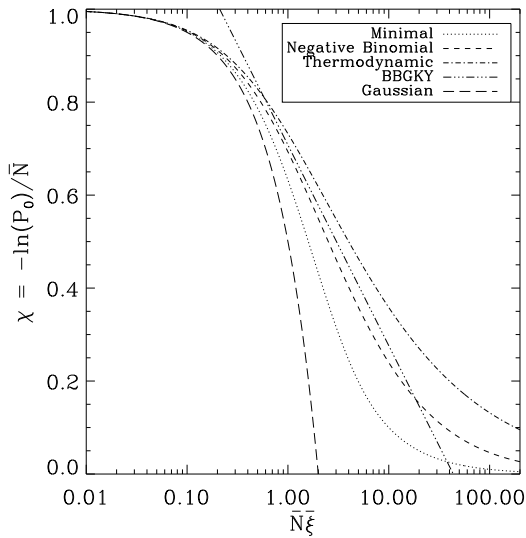}
\caption{Reduced VPF for several of the hierarchical models discussed
in $\S$\ref{s:thy}.  The models are most easily distinguished from one
another for $\bar{N}\bar{\xi}>5$,  a region that can be investigated
with currently existing low- and high-redshift galaxy samples.  Note
that the Poisson model requires $\chi=1$.}\label{fig:models}
\end{figure}

As the dynamical equations governing gravitational clustering can not
be solved in the weakly to strongly nonlinear regime using
perturbation theory (or any of its offspring), various
phenomenological models have been proposed to relate the higher order
correlation functions to the two-point correlation function.  These
models provide a complete description of gravitational clustering, yet
each is inadequate either theoretically or observationally.  See
Fig.~\ref{fig:models} for examples of predictions of the VPF from some
of the more popular models; for a detailed treatment see \citet{Fry86,
Fry88}.

In a Poisson model, all moments with $p>1$ vanish, and the VPF can be
described simply by:
\begin{equation}
P_0 = e^{-\bar{N}}, \,\,\,\,\,\,\,\,\,\, \chi=1.
\end{equation}
A Gaussian model is almost as simple and equally inapplicable to the
statistics of large scale structure in the universe on nonlinear
scales at late times.   This model implies that the entire statistical
distribution is described by the two-point moment (in this case
$\bar{\xi}$) and that all higher moments identically vanish.  The
Gaussian and Poisson models are obviously  not actually hierarchical
models; we include them under 'hierarchical models'  for simplicity.

More realistic models rely on the hierarchical \emph{Ansatz} to
specify all possible moments of the galaxy distribution.  These models
can be thought of as prescriptions for the $S_p$ scaling coefficients.
The simplest model in this case forces $S_p=1$ for all $p$ and is
sensibly called the `Minimal' model \citep{Fry86}.  Yet another model
was motivated by an attempt to combine the hierarchical \emph{Ansatz}
with the BBGKY kinetic equations \citep{Fry84a}.  The BBGKY model
becomes unphysical for large void radii, as can be seen in
Fig.~\ref{fig:models} where $\chi$ becomes negative for
$\bar{N}\bar{\xi}>30$.

The `thermodynamic model' was first proposed by \citet{Saslaw84} and
arose from a theory which treated galaxy clustering by analogy to
statistical mechanics.  It was later extended  into a more
self-consistent model by \citet{Fry86}:
\begin{equation}\label{eqn:thermo}
\chi = [(1+2\bar{N}\bar{\xi})^{1/2}-1]/\bar{N}\bar{\xi},
\end{equation}
\begin{equation}
\label{eqn:s_thermo}
S_p = (2p-3)!!.
\end{equation}

Intriguingly, a model developed by \citet{Sheth98} which combines an
excursion set  approach to the evolution of the halo mass function
with a simple model  for the spatial distribution of such halos
predicts the same values for $S_p$.  \citet{Fry85} points out that
this theory is only strictly true for large volumes, and that the
derivations do not apply for $\bar{\xi}\sim1$ ($r\sim5$ $h^{-1}$ Mpc).
It is also difficult to understand how such large scales could have
become thermodynamically relaxed over the age of the universe.
\citet{Hamilton85} compared this model prediction to the VPF from the
CfA survey, but their results were inconclusive.  They did find,
however, that volume limited subsamples of different minimum
luminosities  were hierarchically related in that samples with
different $\bar{N}$ and $\bar{\xi}$, had similar $S_p$.

The final model we consider has had a colorful history.  The negative
binomial distribution, also known as the modified Bose-Einstein
distribution, was used early on by \citet{Carruthers83a} to study the
count distribution of charged hadrons resulting from high-energy
collisions and by \citet{Carruthers83b} to describe the observed
distribution of Zwicky clusters.   The model can be described by:
\begin{equation}
\label{eqn:nb}
\chi = ln(1+\bar{N}\bar{\xi})/\bar{N}\bar{\xi},
\end{equation}
\begin{equation}
\label{eqn:s_nb}
S_p = (p-1)!,
\end{equation}
\begin{equation}
\label{eqn:nbvpf}
P_0 = \left[\frac{1}{1+\bar{N}\bar{\xi}}\right]^{1/\bar{\xi}}.
\end{equation}

\citet{Gaztanaga93} analyzed the reduced  VPF in the SSRS2 and CfA
redshift surveys but could not discriminate between  the thermodynamic
and negative binomial models because of the size of their errors.  The
negative binomial distribution was re-derived in their appendix by
considering a sample divided into small cells, with the occupation
probability  of each cell depending only on $\bar{\xi}$, and being
independent of the  other cells \citep[see also][]{Elizalde92}.
Unfortunately, the derivation contains little insight into the
physical mechanisms that might drive point  distributions to become
negative binomial.  This model has also been derived from
thermodynamic arguments \citep{Sheth95}.

\citet{Mo90} and \citet{Vogeley91}  independently analyzed the CfA
survey and found the data to be more consistent with the negative
binomial than thermodynamic model over a range of luminosity
thresholds and morphological types, though the agreement between model
and data was not conclusive.  Most recently, \citet{Croton04} found
that galaxies in the 2dF survey follow the negative binomial model
over a range of differential luminosity bins.

\subsection{Convergence Issues}\label{sec:conv}

The astute reader will have noticed two problems with the theoretical
models presented above, one related to convergence and the other to
the unintuitive nature of the sums in Eqns.~\ref{eqn:vpf},
\ref{eqn:fry}  and \ref{eqn:chi} .  The form for the $S_p$ values
associated with the thermodynamic and negative binomial models
(Eqns.~\ref{eqn:s_thermo} and \ref{eqn:s_nb})  can be understood as
arising from a Taylor series expansion for the given  $\chi$ values in
Eqns.~\ref{eqn:thermo} and  \ref{eqn:nb}.  Such an expansion is,
however, only valid for $\bar{N}\bar{\xi}<1$; the sum rapidly diverges
if $\bar{N}\bar{\xi}>1$.  The conclusion must be that the $S_p$ values
often quoted for these models cannot be correct for large
$\bar{N}\bar{\xi}$, i.e. for $r\gtrsim3$ $h^{-1}$ Mpc (the  precise
$r$ at which $\bar{N}\bar{\xi}=1$ depends on the sample).

Realizing this issue, we have explored other forms for the $S_p$
values such that the sums in Eqns.~\ref{eqn:fry} and~\ref{eqn:chi} do
converge. Simplistic models such as:
\begin{equation}
\label{eqn:sp}
S_p = e^{Ap+B},
\end{equation}
where A and B are free parameters, are consistent with recent
observations of $S_p$ for $p<6$ \citep{Baugh04} and are also
consistent with $(p-1)!$ for small $p$.  Although the sums converge
with these $S_p$ values, they do not converge to a hierarchical model;
in fact, for many values of A and B, the sums converge to non-physical
values ($P_0<0$).  It is unclear how far one should  pursue this,
given (1) that the $S_p$ values are probably scale-dependent, and here
we have assumed that they are not, (2) there is no theoretical reason
to suspect that the  higher-order moments should be simply
proportional to a power of  $\xi$, and (3) though mathematically
valid, the expansion  of $P_0$ in terms of $\xi_p$ is not terribly
useful in practice.

To illustrate (3) consider the following.  One might think that
increasingly higher-order moments would be increasingly {\it
irrelevant} for void statistics.  In fact, we have explicitly computed
the sum in Eqn.~\ref{eqn:vpf} for the first eight correlation
functions in  large galaxy mock catalogs and find that the resulting
partial sum in no way approximates the VPF measured from the same mock
catalogs.  It seems that increasingly higher-order moments are, in
fact,  increasingly {\it relevant}.

In order to gain a clearer understanding of this we have investigated
the sequence of partial sums of Eqn.~\ref{eqn:vpf} using the $S_p$
values given in Eqn.~\ref{eqn:sp}.  We find that the sequence of
partial sums for $P_0$ oscillates wildly for $p<20$, but rapidly
converges for $p>25$.  In other words, it is not the first few, but
the first $\sim25$ correlation functions that are necessary to
accurately describe the void distribution.  Though it is comforting to
know that the sum eventually converges, it is quite puzzling why
convergence should require so many correlation functions.

One possible explanation for the importance of higher-order moments is
the following: since there exist clusters with twenty objects or more,
we might  expect the 20-point correlation function to be non-zero.
This function describes the probability, in excess of random, that a
region of space  which contains 19 objects will contain a 20th.  Since
this will almost  never happen for a random sample, the 20-point
function will likely be quite large, at least  on small scales.  Hence
one might expect contributions from $p$-point  correlation functions
as long as there are clusters of objects with $p$ members.

\section{Methodology}\label{s:meth}
We next describe our general methodology for obtaining void statistics
from large samples of galaxies.  Issues concerning specific survey
details,  such as the proper handling of angular window functions, are
treated as they arise in $\S\S$ 5 and 6, where we use data from
different surveys.

\begin{figure}
\plotone{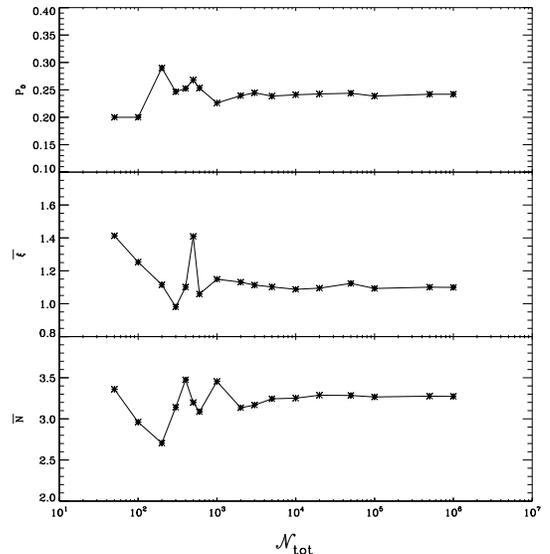}
\caption{Convergence of measured quantities as a function of the
number of  randomly placed spheres ($\mathcal{N}_{tot}$) for galaxies
in our simulations  with a sphere radius of $7$ $h^{-1}$ Mpc.  The
rapid convergence is due to the fact that the  simulation volume is
well-sampled once the number of test volumes  exhausts the number of
independent configurations, which for this simulation, at a test
sphere of $R=7$ $h^{-1}$ Mpc,  occurs at $\mathcal{N}_{tot}\approx
1500$.}
\label{fig:convergence}
\end{figure}

Investigating the VPF and reduced VPF requires the measurement of
three quantities: $\bar{N}$, $\bar{\xi}$, and $P_0$, all of which are
functions of the sphere radius, $R$.  Each of these quantities is
straightforwardly determined by a counts-in-cells (CIC) approach.  One
simply places large numbers ($\sim10^5$) of random spheres within the
survey and counts the number of galaxies contained within each sphere.
This is then repeated for many sphere radii.  $\bar{N}$ is the average
number of galaxies in a sphere:
\begin{equation}
\bar{N} = \frac{1}{\mathcal{N}_{tot}}\sum_{i=1}^{N_{tot}}N_i
\end{equation}
where $\mathcal{N}_{tot}$\footnote{Here and throughout we use
$\mathcal{N}$ to  refer to the number of spheres and $N$ to refer to
the number of galaxies.} is  the total number of spheres placed and
$N_i$ is the number of galaxies in the $i$th sphere.  $P_0$ is the
number of spheres containing zero galaxies divided by the total number
of spheres,
\begin{equation}
P_0 = \frac{\mathcal{N}_0}{\mathcal{N}_{tot}}
\end{equation}
where $\mathcal{N}_0$ is the number of spheres that contain zero
galaxies, and $\bar{\xi}$ is the variance in the number of galaxies
per sphere:
\begin{equation}
\bar{\xi} = \frac{\overline{(N-\bar{N})^2} - \bar{N}}{\bar{N}^2}
\end{equation}
In the limit of large numbers of random points, the CIC approach for
determining $\bar{\xi}$ is known to be mathematically equivalent to
more conventional methods \citep{Szapudi98}.  We independently confirm
this result by comparing our CIC-measured $\bar{\xi}$ to the
volume-averaged correlation function obtained via the popular
Landy-Szalay estimator \citep{Landy93}.  We find that the two
approaches are entirely consistent.  To test the sensitivity  of these
measured quantities to the total number of spheres used,
$\mathcal{N}_{tot}$,  we have computed $P_0$, $\bar{N}$, and
$\bar{\xi}$ using a sphere  radius of $7$ $h^{-1}$ Mpc, as a function
of $\mathcal{N}_{tot}$ (Fig.~\ref{fig:convergence}),  for mock
galaxies in a simulation box of length $256$ $h^{-1}$ Mpc (see below
for simulation details).  As expected, for small $\mathcal{N}_{tot}$,
these quantities  are unstable, but for $\mathcal{N}_{tot}\gtrsim10^4$
the quantities converge well.   This rapid convergence holds for both
smaller and larger void radii and is  due to the fact that the
simulation volume is well-sampled once the  number of test volumes
exhausts the number of independent configurations.   To be
conservative, we use $\mathcal{N}_{tot}\sim10^5$ to calculate these
quantities.  Errors for all measured quantities are estimated by
``jackknife'' sampling.

Before moving on, we should highlight the fact that all quantities
used  in analyses of void statistics are spherically averaged.  This
type of averaging effectively throws away much  of the useful
information contained  in higher order statistics.  For example,
$\xi_3$ measures the skewness of the  distribution, and will in
general {\it not} be spherically symmetric.  Hence it  is unclear
whether or not spherically averaging $\xi_3$ is even the proper way to
average $\xi_3$; these ambiguities make interpretation difficult.

The $\bar{\xi}$ used here and in other void analyses refers to the
correlation function in \emph{redshift space}; the correlation
function has not been de-projected into real space, which is the more
conventional way of reporting $\xi$.  This is important because, as we
will see, the real space VPF measured in mock catalogs is different
from the behavior of the VPF in redshift space, and our main
conclusions turn out to be valid only in redshift space.  Indeed,
\citet{Vogeley94} find that, at fixed void radii, $P_0$ in simulations
is larger in redshift space compared to real space, with the
difference  increasing at larger void radii.  These authors also find
that redshift  space void statistics closely follows one or another
hierarchical model  (which model the statistics follow depends on
gross cosmological models,  i.e. closed v.s open and biased
vs. unbiased),  while the real space void statistics do not.  One
example of redshift-space effects relevant to this study is the
flattening of the $S_p$ scaling coefficients in redshift space when
compared to real space.  This is due to the reduction of clustering
strength on small scales caused by peculiar velocities (known as the
fingers-of-God effect).  See \citet{Bernardeau02} for examples of this
and other redshift space effects.

Finally, we require that the number density of the sample under
consideration be independent of redshift, and so construct
volume-limited samples for our analysis.  Specifically, we require
that each galaxy be  observable over the entire redshift range we
consider.

\section{Void Statistics in Simulations} \label{s:mocks}

In this section we analyze void statistics in mock galaxy catalogs
constructed from N-body $\Lambda$CDM simulations in order to help
understand and interpret void statistics recovered from observational
samples.  First we measure  the VPF and reduced VPF for dark matter
particles and mock galaxies at  $z\sim0$ and $z\sim1$.  We then
investigate the reduced VPF for different  halo model halo occupation
distributions (HODs) and find that they  are similar over a wide range
of  HOD parameters.  Next, we determine the effects of redshift space
distortions  on the VPF and finally we compare VPFs for halo centers
alone to VPFs for  mock galaxies.  Throughout this section we use full
simulation boxes at  two outputs: $z=0.087$ (``$z\sim0$'') and
$z=0.92$ (``$z\sim1$'');  in $\S$\ref{s:deep}, where we investigate
the effects of survey geometry  on the VPF, we extract light-cone
geometries from the full simulation.  Except for $\S$\ref{s:rse}, all
analyses in this section are performed for  galaxies, dark matter
particles, and dark matter halos, in redshift space.

\subsection{The Simulations}\label{s:sims}

The mock galaxy catalogs we use were constructed specifically for the
DEEP2 survey.  A complete description of the catalogs is given in
\citet{Yan04}; we give the relevant details here.  N-body simulations
of $512^3$ dark matter particles with a particle mass
$m_{part}=1.0\times10^{10}$ $h^{-1} M_{\Sun}$ were run in a
$\Lambda$CDM universe using the TreePM code \citep{MWhite02} in a
periodic, cubical box of side length $256$ $h^{-1}$ Mpc.  Dark matter
halos were identified by running a ``friends-of-friends'' group
finder.  Galaxies were then inserted via a halo model approach, where
galaxies are placed in dark matter halos using a simple prescription
\citep[see e.g.][]{Seljak00,Ma00,Peacock00,Cooray02,Kravtsov04}.

The main ingredient of the halo model is the halo occupation
distribution,  HOD, which specifies the mean number of galaxies to be
placed in a given halo, as a function of the halo mass ($\langle
N(M)\rangle$).  In its most common form the HOD is the sum of a
power-law describing the sub-halo (satellite) population and a step
function above some minimum mass, which describes the host halo
(central galaxy) \citep{Kravtsov04}.  Luminosities are then assigned
to galaxies in a halo according to a conditional luminosity function
(CLF), $\Phi(L|M)$, which specifies the luminosity function of
galaxies in halos of mass $M$ \citep{Yang03}.  The CLF is
traditionally chosen to have a Schechter form.  The \emph{same} HOD
and CLF are used to populate halos at both high and low redshift, with
only the underlying dark matter distribution and the characteristic
scale of the luminosity function, L$^*$, evolving with redshift
(evolution in L$^*$ was taken  to be 1 mag based on COMBO-17 data
\citep{Wolf03}).  This ``no evolution''  hypothesis produces catalogs
that are in agreement with two-point clustering measurements  from
DEEP2 at $z\sim1$ and the $B_j$-band luminosity function and two-point
clustering of  2dF galaxies at $z\sim0$ \citep{Madgwick03c}.  For what
we call  the `primary' simulations, the specific parameters of the HOD
and CLF are chosen to best match the two-point clustering measurements
at $z\sim1$.

\subsection{VPF of Dark Matter and Mock Galaxies}\label{s:sim_res}

\begin{figure}
\plotone{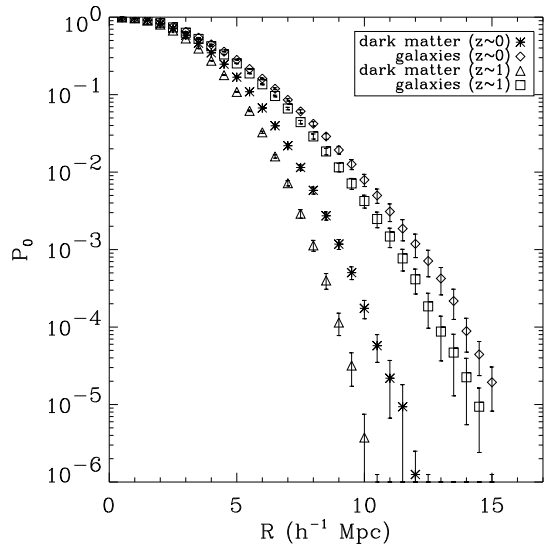}
\caption{VPF ($P_0$) measured in mock galaxy catalogs and dark matter
N-body  simulations as a function of comoving radius at $z\sim1$ and
$z\sim0$.  It is apparent that  dark matter at $z\sim0$ (asterisks)
has larger voids, and hence is more clustered, than  dark matter at
$z\sim1$ (triangles).  Similarly, galaxies at $z\sim0$ (diamonds) are
more clustered than galaxies at $z\sim1$ (squares), and galaxies have
more voids than dark matter particles at both epochs..  All samples
shown  here were randomly diluted to a number density of $n=0.009$
$h^{3}$Mpc$^{-3}$, and  the mock galaxies were chosen to have
luminosities in the range
$-22<M_B-5\rm{log}(h)<-19$.}\label{fig:mocks_vpf}
\end{figure}

This section uses the `primary' simulations as described above.  All
samples drawn from simulations have been randomly diluted to the same
number density ($n=0.009$ $h^{3}$ Mpc$^{-3}$) to allow for a direct
comparison between VPF measurements (recall that the VPF, as opposed
to the  reduced VPF, is quite sensitive to $\bar{N}$).  Furthermore,
all galaxies  in these samples are restricted to have luminosities
$-22<M_B-5\rm{log}(h)<-19$.  Fig.~\ref{fig:mocks_vpf} shows the VPF
calculated for dark matter particles and galaxies at $z\sim1$ and
$z\sim0$ in redshift space,  where errors are estimated using
jackknife sampling.  There are two obvious trends.  The first is that,
at both redshifts, galaxies have a larger $P_0$ at all radii compared
to dark matter, indicating that they are more highly clustered and
that galaxies included here are biased relative to the dark matter.
This simply results from the fact that these galaxies live in massive
dark matter halos that are large overdensities and hence are more
highly clustered than the overall dark matter distribution.  This is
not in conflict with  the realization that at $z\sim0$ the galaxy
distribution accurately traces the dark matter  on large scales
\citep{Verde02} because (1) we are probing different scales and  (2)
these studies have only measured the linear and quadratic bias terms,
whereas  we are in principle sensitive to all non-linear biasing
terms.   The second result in Fig.~\ref{fig:mocks_vpf} is that both
galaxies  and dark matter become more clustered with time, as seen by
the larger  voids at $z\sim0$ relative to $z\sim1$; this is due to the
effects of gravity.

\begin{figure}
\plotone{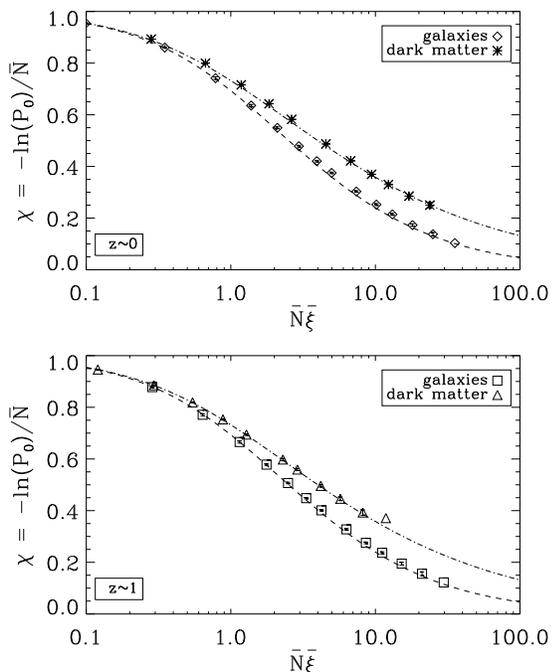}
\caption{Reduced VPF measured in redshift space from  simulations at
$z\sim0$ (\emph{top}) and $z\sim1$ (\emph{bottom}).  At both redshifts
the galaxies follow the negative binomial model (dashed line) while
the dark matter follows the thermodynamic model (dot-dashed line).
Since all samples  have been diluted to the same number density, this
implies that the trends seen in Fig.~\ref{fig:mocks_vpf} are due
solely to  changes in the volume-averaged two-point correlation
function.  See text for details.}
\label{fig:mocks_rvpf}
\end{figure}

In Fig.~\ref{fig:mocks_rvpf} we show the reduced VPF separately for
low- and high-redshift mock samples as well as for two hierarchical
models, the negative binomial (dashed) and thermodynamic (dash-dot).
Here as elsewhere,  the upper limit on $\bar{N}\bar{\xi}$ in the
reduced VPF is set  by the radius at which $P_0=0$; at that point
$\chi$ becomes undefined.  The error on the mean in each point
(computed by dividing the cube into  octants and measuring void
statistics within each octant) is smaller than the plotted symbols
($\sim0.001$), although errors are highly covariant  between different
sphere radii.  In principle there should be errors  on
$\bar{N}\bar{\xi}$ in addition to $\chi$, as $\bar{N}\bar{\xi}$ will
vary across the octants at a specific $R$.  In practice however,  the
error on $\bar{N}\bar{\xi}$ is negligible as it is smaller  than the
binsize.

The dark matter VPF is well described by the thermodynamic model at
all scales for both $z\sim1$ and $z\sim0$, while  the galaxy VPF fits
the negative binomial model to a precision better than the  achievable
observational errors.  In fact, the dark matter matches the
thermodynamic model well to at least $z\sim2$, the highest redshift
simulation output available.  This directly shows that the VPF can be
entirely parameterized in terms of two variables, $\bar{N}$ and
$\bar{\xi}$, since, for a given hierarchical model, $\chi$ is a
function of only of their product  (see Eqn.~\ref{eqn:nbvpf}).  The
differences seen in the VPF for galaxies at low and high redshift
(Fig.~\ref{fig:mocks_vpf}) are described entirely by the evolution of
$\bar{\xi}$ with cosmic time.  Below we find that this dependence
holds in the data as well, not only as a function of  redshift but
also for samples with varying galaxy properties at the same redshifts.
\citet{Croton04} do not find this good agreement between the
thermodynamic  model and dark matter particles drawn from the Hubble
Volume simulations.   We speculate that the much larger particle mass
($m_{part}\sim 10^{12}$ $h^{-1} M_\Sun$)  of these simulations does
not afford an accurate measurement of the dark matter  particle
reduced VPF.

There is a small but perceptible up-turn in the reduced VPF for all
samples in Fig.~\ref{fig:mocks_rvpf} at the largest $\bar{N}\bar{\xi}$
measured.  A much larger up-turn has been seen in the CfA survey
\citep{Vogeley91, Vogeley94} at similar void radii ($7-15$ $h^{-1}$
Mpc, depending  on the luminosity threshold).  This may be due to
fluctuations in the initial conditions of the simulations, as both the
galaxies and dark matter at low and high redshift were derived from
the same simulation.  Furthermore, these large scales (corresponding
to void diameters of $\sim30$ $h^{-1}$ Mpc) are approaching the limit
at which a simulation with box size of $L=256^3$ $h^{-3}$ Mpc$^3$
becomes unreliable due  to the fact that large-scale modes, which
become increasingly important on  larger scales, cannot be included in
simulations with periodic  boundary conditions.  Future studies with
different realizations of the initial conditions and larger box sizes
will be able to address the significance of this
deviation. Regardless, we are unable to probe these largest void
scales with the currently available galaxy surveys.

Finally, we find that the good agreement between the reduced VPF in
simulations and  the negative binomial model is insensitive to random
dilutions of the  mock galaxies.  Specifically, we randomly dilute the
mock galaxy catalogs to $25$\%, $50$\%, and $75$\%  of the nominal
number density and find no differences in the resulting reduced VPFs,
as expected  since removing dependencies on $\bar{N}$ was a major
reason for introducing the reduced VPF  in the first place.  Defining
the VPF as a sum of higher  order correlation functions and then
inserting the hierarchical \emph{Ansatz}  provides a simple
explanation for this insensitivity of the reduced VPF on  $\bar{N}$.

\subsection{Constraining the Halo Model} \label{s:constrain}

As mentioned above, the halo model can be very useful for placing
galaxies in cosmological dark matter simulations.  It also provides a
way of understanding the VPF without resorting to an infinite sum of
higher-order correlation functions.  Specifically,
\begin{equation}
P_0(V) = \int_0^{\infty}
P(M<M_{min}|\delta_{\bar{m}})P(\delta_{\bar{m}})d\delta_{\bar{m}},
\end{equation}
where $\delta_{\bar{m}}$ is the mass density contrast smoothed over a
volume $V$, $P(\delta_{\bar{m}})$ is the probability of having a given
smoothed mass density contrast, and $P(M<M_{min}|\delta_{\bar{m}})$ is
the probability of finding a halo with mass less than $M_{min}$ in a
volume $V$ with smoothed density contrast $\delta_{\bar{m}}$.  Here we
have written $P_0(V)$ explicitly as a function of the smoothing
volume, but note that it is the same $P_0$ that occurs elsewhere.  The
only input from the halo model is the value for $M_{min}$, the mass
above which halos  contribute to $P_0$.  $P(\delta_{\bar{m}})$ is
variously assumed in the literature to have a lognormal, negative
binomial, or Gaussian distribution, and
$P(M<M_{min}|\delta_{\bar{m}})$ is fixed by the relation between halos
and dark matter (the halo bias), which can be determined from
simulations.

We would like to test the ability of the VPF to   constrain the halo
model, as suggested by \citet{Berlind02}.  Their main results are (1)
that the VPF is sensitive to M$_{min}$, the minimum mass that a dark
matter halo must have to host a galaxy, and (2) that the VPF is
entirely insensitive to the spatial distribution of galaxies within
halos.

\begin{figure}
\plotone{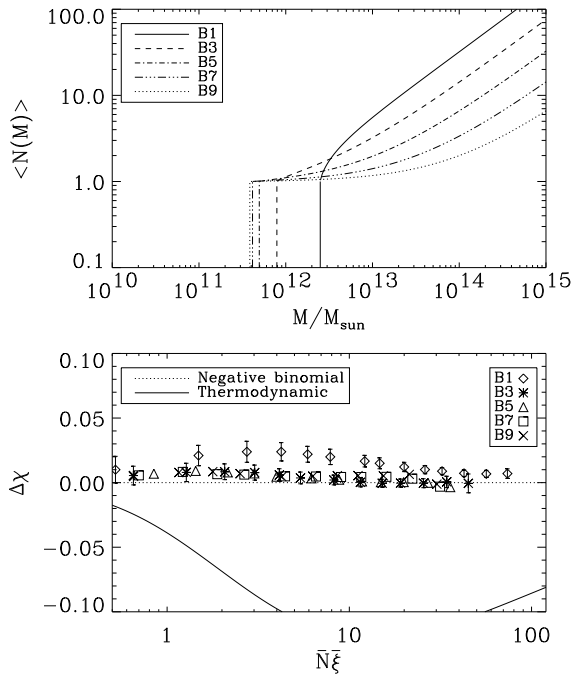}
\caption{\emph{Top}: Various HODs used to populate galaxies in  dark
matter halos in simulations.  \emph{Bottom}: The resulting reduced
VPF, in redshift  space, for the simulations with the HODs shown in
the top panel, plotted as  differences from the negative binomial
model.  The only HOD inconsistent with  the negative binomial model,
model B1, is easily ruled out observationally by its two-point
correlation  function.  The reduced VPF is almost entirely insensitive
to the number of  galaxies placed in a halo as a function of galaxy
mass.  Furthermore, any changes seen in the  (un-reduced) VPF can be
solely attributed to changes in $\bar{\xi}$.  Hence studying  the VPF
seems to provide no new constraints on the HOD beyond the information
available  in $\bar{\xi}$.}\label{fig:HOD}
\end{figure}

To investigate these claims we calculate void statistics,  in redshift
space, for mock galaxy catalogs constructed from various HODs.  The
top panel in Fig.~\ref{fig:HOD} shows five different HODs with the
simple form \citep{Kravtsov04}:
\begin{displaymath}\label{eqn:hod}
\langle N(M)\rangle = \left\{ \begin{array}{ll} 1 +
\left[\frac{M-M_{min}}{M_1}\right]^{\alpha} & \textrm{for
$M>M_{min}$}\\ 0 & \textrm{for $M<M_{min}$}
\end{array} \right.
\end{displaymath}
where $\alpha=0.75$ and M$_{min}=25.0, 7.9, 4.9, 4.1$ and $3.9$ (in
units of  $10^{11}$ $h^{-1}$ M$_{\Sun}$).   For each M$_{min}$, M$_1$
is determined by  fixing the overall number density at 0.01 $h^{3}$
Mpc$^{-3}$.  These five  models are referred to as B1, B3, B5 ,B7, and
B9, respectively.

In order to clearly show small variations between the models, in
Fig.~\ref{fig:HOD} (bottom panel) we plot differences between the
negative binomial model and the measured reduced VPF for each galaxy
catalog.   The reduced VPF remains essentially unchanged for these
different HODs,  leading us to conclude that the VPF can be entirely
determined by $\bar{\xi}$, i.e., by Eqn.~\ref{eqn:nbvpf}.  The scatter
between the five models is $\sim0.02$, much smaller than the
attainable accuracy with even the largest current galaxy surveys; we
therefore conclude that over a wide range of physically meaningful
HODs, the VPF cannot constrain M$_{min}$ beyond constraints attainable
from $\bar{\xi}$ alone.  In the figure, the reduced VPF for the B1 HOD
is the most deviant from the  negative binomial model.  However,
observations of the two-point correlation function  at $z\sim1$
\citep{Coil05a} rule this model out.

These results are consistent with the claims of \citet{Berlind02},  as
they were concerned with void statistics in \emph{real space}.   The
halo model formalism becomes much more complex analytically in
redshift space, due to the presence of peculiar velocities.   We find
using simulations, that redshift-space effects essentially  `wash out'
any differences in the VPF due to different HODs.  Qualitatively,  one
would expect that placing more galaxies in halos would increase the
$S_p$ values on  small scales in real space.   However, in redshift
space, small-scale  power is more smeared out because there are now
more galaxies residing  in massive halos with higher velocity
dispersions.  It is only a surprise that these two effects should act
to cancel each other out exactly.  We now investigate the effects of
redshift-space distortions  on void statistics in detail.

\subsubsection{Redshift Space Effects}\label{s:rse}

The effects of redshift space distortions on clustering statistics
have been studied  in detail \citep[see e.g.][]{Lahav93, MWhite01,
Seljak01, Scoccimarro04},  and have been probed with void statistics
\citep{Vogeley94, Ryden96}.  \citet{Vogeley94} found that voids appear
somewhat larger in redshift space  and that the reduced VPF in real
space does not agree with any of the  hierarchical models.

We explore the extent to which redshift-space effects wash away
information by  calculating the VPF resulting from several HODs with
fixed $M_{min}$ and varying  $\alpha$ (for $0.5<\alpha<1.0$),
i.e. varying the number of galaxies placed in  halos with mass above
$M_{min}$, while still keeping the overall number density fixed  at
0.01 $h^{3}$ Mpc$^{-3}$.  One might imagine that, by fixing $M_{min}$
and varying  $\alpha$, $P_0$ would remain unchanged, but $\bar{\xi}$
would vary because  adding more galaxies to a halo will strongly
affect the small scale clustering.  If this were true, then $P_0$
would not simply be a function of $\bar{\xi}$  alone.  By analyzing a
suite of mock galaxy catalogs constructed with  these various HODs, we
find, as expected, that $P_0$ remains unchanged.  Yet surprisingly,
but in accordance with  the idea that redshift-space distortions
conspire to `wash out' differences  due to different HODs, $\bar{\xi}$
remains unchanged for $R\gtrsim1$ $h^{-1}$ Mpc  and is only mildly
different for $R\lesssim1$ Mpc.  By showing that $\bar{\xi}$ remains
the same across different HODs when $P_0$ and $\bar{N}$ remain the
same, we further solidify our claim that $P_0$ is solely a function of
$\bar{N}\bar{xi}$,  and hence that $P_0$ does not contain additional
information beyond what is  available in $\bar{\xi}$.

\begin{figure}
\plotone{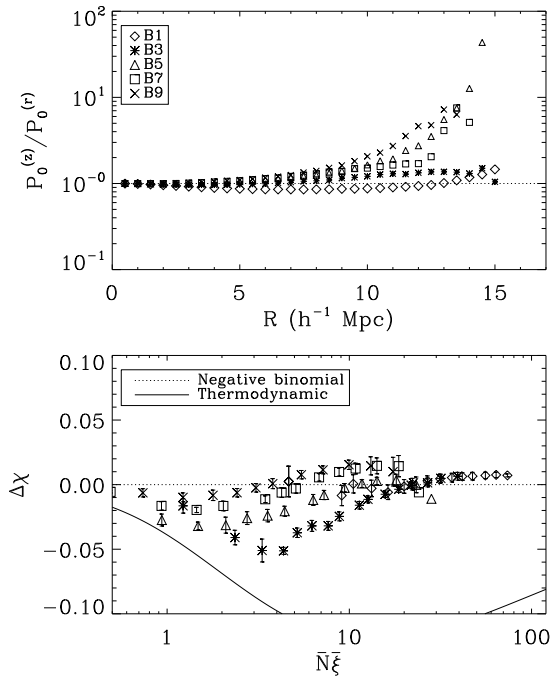}
\caption{Effects of calculating void statistics in real space
vs. redshift space.   \emph{Top}: Ratio of the VPF, $P_0$, in redshift
space to real space.   There are more voids on large scales in
redshift space because of coherent galaxy  infall into groups and
filaments.  \emph{Bottom}: Reduced VPF in real space for mock  galaxy
catalogs, plotted as differences between the mock galaxies and the
negative  binomial model.  Unlike the reduced VPF in redshift space,
here there are  trends for different HOD parameters; smaller values of
$M_{min}$ agree less well  with the negative binomial model than
larger values.}\label{fig:HOD_rz}
\end{figure}

To quantify the differences between real and redshift space we
calculate void statistics  for our five HOD models (B1-B9) in real
space and compare them to their redshift space analogs.  The  top
panel in Fig.~\ref{fig:HOD_rz} displays the fractional probability
increase  of the VPF when comparing redshift space ($P_0(z)$) to real
space ($P_0(r)$).   Two competing effects are at work here: (1)
peculiar velocities smear out small scale clustering, and (2) Kaiser
infall, due to the coherent infall of  structures on larger scales
\citep{Kaiser87}, which has the effect of increasing the size of voids
on large scales in redshift space.

The general trends in this figure are consistent with the underlying
HOD of  each sample.  For example, model B1 preferentially places
galaxies in  higher-mass halos compared to other HODs, and more
massive halos have larger  velocity dispersions.  The larger
dispersions   decrease the size of voids as seen in redshift space,
explaining why there  are fewer voids in redshift space compared to
real space for the B1 model.   This model also has the smallest
differences on large scales.  This is in part because  velocity
dispersions are small compared to the largest voids, but also because
of Kaiser infall, which increases the apparent size of voids in
redshift space on large scales,  is important primarily for lower mass
halos that are falling into forming structures.   Since $M_{min}$ is
so large in the B1 model, we expect that most halos in this model  are
not falling onto forming superstructures.

In the bottom panel of Fig.~\ref{fig:HOD_rz} we see that the reduced
VPF can in fact uniquely  constrain the halo model when the analysis
is performed in real space.   The VPF for a  set of catalogs can be
entirely described by a single function  $f(\bar{\xi},\bar{N})$, and
hence cannot be used to uniquely constrain models, if the reduced VPF
for those models are all the same.  Yet for our five  HOD models, this
is not the case in real space.  Instead, there is a  definite trend of
$\chi$ with $M_{min}$; samples with smaller $M_{min}$ are less like
the negative binomial model.  On larger scales ($\sim5$ $h^{-1}$ Mpc)
the HODs are again indistinguishable, as is expected since HOD
parameters are  relevant only on halo scales  (i.e. $\sim1$ $h^{-1}$
Mpc).  Although it seems intuitive that the VPF should be capable  of
uniquely constraining the HOD (specifically $M_{min}$), redshift-space
effects  make this extremely difficult in practice.

\subsubsection{VPF for Dark Matter Halos}

\begin{figure}
\plotone{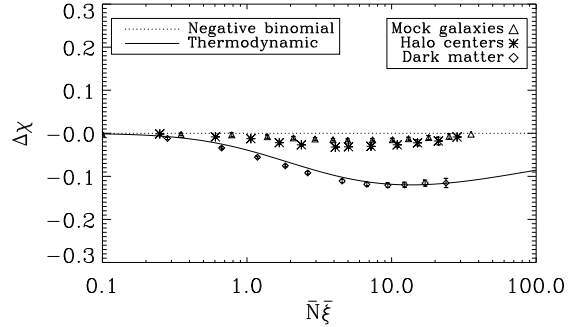}
\caption{Reduced VPF for mock galaxies, dark matter halo centers, and
dark matter  particles measured in redshift space from the `primary'
simulation, plotted as differences from  the negative binomial model.
It is primarily the clustering properties  of dark matter halos, not
the galaxies within them, that generates  agreement with the negative
binomial model.  }\label{fig:halo}
\end{figure}

We can better understand the insensitivity of the VPF to particular
parameters  of the HOD by considering the VPF for the {\it centers} of
dark matter halos  (restricted to have $M_{halo}>M_{min}$) alone.
Fig.~\ref{fig:halo} compares  the reduced VPF for dark matter
particles, dark matter halo centers, and  mock galaxies.  As the
figure indicates, it is primarily the  clustering properties of dark
matter halos, not the number of galaxies  within them, that generates
agreement with the negative binomial model.

We find the similarity in reduced VPF for halo centers and mock
galaxies intuitive  for two reasons.  First, as most dark matter halos
in our simulations with  $M>M_{min}$ contain a single isolated galaxy,
and galaxies in most halo models are placed first in the center of the
halo, it is reasonable that the clustering of  halo centers has more
impact on the galaxy reduced VPF than $\alpha$.  Second,  when more
than one galaxy is placed in a halo, it is typically within $\sim500$
$h^{-1}$ kpc of the halo center, as very few halos have larger radii.
Hence  a sphere containing a satellite galaxy will typically already
include the galaxy  at the halo's center.  This also explains why void
statistics are insensitive  to the spatial distribution of galaxies
within halos, as noted by \citet{Berlind02}.

In conclusion, the VPF does not seem to be capable of providing
competitive constraints on the major parameters of the HOD.  This is
due both to redshift space distortions washing out information on
small scales  and to the fact that the clustering of dark matter halos
alone dominates the VPF.  Although one might have thought that the VPF
for galaxies and halos would be uniquely sensitive to $M_{min}$, it
turns out that any changes in $M_{min}$ are likewise reflected in
$\bar{\xi}$, in a way that can exactly account for changes in the VPF.

\section{Void Statistics from DEEP2}\label{s:deep}
We now measure void statistics in state-of-the-art redshift surveys
that are mapping the  three-dimensional positions of thousands of
galaxies.  In this section we probe void statistics when the universe
was roughly half its present age ($z\sim1$).   In the following
section we focus on voids in the local universe ($z\sim0$).

\subsection{The Data}
The DEEP2 Galaxy Redshift Survey \citep{Davis04} is an ongoing project
that is gathering optical spectra for $\sim50,000$ galaxies at
$z\sim1$ using the DEIMOS spectrograph on the Keck II 10-m telescope.
The completed survey will span a comoving volume of $\sim10^6 h^{-3}$
Mpc$^3$, covering $3$ deg$^2$ over four widely separated fields;
observations began  July 2002 and are expected to be completed in
mid-2006.  Due to the high dispersion and excellent sky subtraction
provided by the DEIMOS spectrograph  ($R\sim5,000$), our rms redshift
errors, determined from repeated observations, are $\sim30$km
s$^{-1}$, minuscule compared to void scales.

Target galaxies are selected using $BRI$ imaging from the CFHT
telescope down to  a limiting magnitude of $R=24.1$ (all magnitudes in
this paper are in the AB system;  see Coil et al. 2004 for photometric
details\nocite{Coil04b}).  In three  of the four fields we also use
apparent colors to exclude objects likely to have $z<0.7$.   This
pre-selection greatly  enhances our efficiency for targeting galaxies
at high redshift \citep{Newman05}.  A fourth field, the Extended Groth
Strip (EGS), has no redshift pre-selection,  and is not used in the
present analysis.  For galaxies with a successfully  identified
redshift, absolute $B$-band magnitudes ($M_B$) and restframe $U-B$
colors,  denoted $(U-B)_0$, have been derived \citep{Willmer05}.  In
the discussion below, we define ``red'' and ``blue'' galaxies
according to whether  $(U-B)_0>1$ or $(U-B)_0<1$; this roughly
corresponds to the saddle point of the color  bimodality observed in
DEEP2 data \citep{Willmer05}.  The current study uses  $12,000$
redshifts with $0.75\gtrsim z\lesssim 1.0$ in three fields covering
$\sim2.2$ deg$^2$.  The three fields we use correspond to the DEEP2
pointings 21, part of 22, 31, 32, 33, 41, and  42, as defined in
\citet{Coil04b}.

An important aspect in analysis of any large-scale galaxy survey is
the  proper handling of selection effects that may vary as a function
of  galaxy property, such as color and luminosity.  DEEP2 is an
$R$-band limited survey,  which corresponds to restframe UV at $z>1$,
and results in fewer red galaxies being targeted at higher redshift
compared to blue galaxies \citep{Willmer05}.  Due to this  selection
effect and the generally lower sampling density beyond $z\sim1$, we
limit  our analysis in this study to $z<1$.

\subsubsection{Survey Geometry and Completeness}\label{sec:deep_compl}
Each DEEP2 field is much longer in the redshift direction than on the
sky; the  $1-2\times0.5$ deg$^2$ fields used for this work span
$40-80\times\sim20$ $h^{-1}$ Mpc  in transverse comoving extent, while
the range $0.7<z<1.0$ corresponds to $560$ $h^{-1}$ Mpc  comoving in
the redshift direction.  The number of possible independent spheres
contained  within the survey, as indicated by $V_{survey}/V_{sphere}$,
varies from $\sim 10^5$  for $1$ $h^{-1}$ Mpc spheres to $\sim 10^3$
for $7$ $h^{-1}$ Mpc spheres.

One might be concerned that the survey geometry  would skew void
statistics when probing voids with diameters comparable to the  short
dimension of the survey.  We have tested this effect by comparing the
VPFs from a full mock galaxy simulation box at $z\sim1$, constructed
from the `primary' simulation  (see $\S$\ref{s:sims}), to mock galaxy
catalogs with a ``lightcone'' geometry similar to DEEP2.  The
lightcone geometry is constructed by stacking together several
simulation box outputs  and slicing through them diagonally, so as not
to pick up the same structures at different times.

\begin{figure}
\plotone{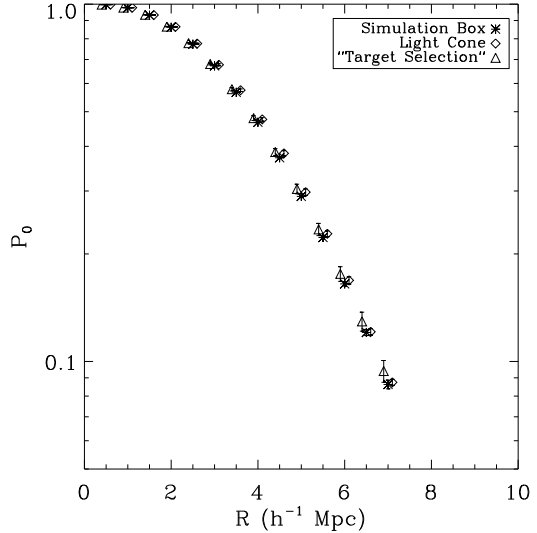}
\caption{VPF, $P_0$, for mock galaxies with $-22<M_B-5\rm{log}(h)<-19$
and a number density of  0.006 $h^{3}$ Mpc$^{-3}$.  $P_0$ values are
for galaxies in the full simulation box at $z\sim1$ (stars), galaxies
in a lightcone with the same geometry as the DEEP2 survey sample in
this paper (diamond), and galaxies in the light cone that have passed
the DEEP2  target selection criteria (triangles).  See the text for
details.  The points have been offset from each other to make the plot
more readable; each set of points is based  on spheres of the same
radius.  Error bars are obtained from averaging over  three mock light
cones, which correspond to the three separate DEEP2
fields.}\label{fig:mock_ccm}
\end{figure}

We find that, even at the largest radii tested , the VPF in the mock
lightcone  and larger simulation box agree to within $1\sigma$
(Fig.~\ref{fig:mock_ccm}).   Of course, due to the decrease in the
number of independent volume elements at  large void radii, cosmic
variance increases, and hence the  scatter (computed across three mock
light cones) increases.   We conclude that the geometry of the DEEP2
survey allows us to accurately  probe voids to radii of at least 7
$h^{-1}$ Mpc comoving.

Another selection effect is the survey completeness.  The DEEP2 survey
spectroscopically targets $\sim$60\% of objects that pass the apparent
magnitude and color cuts mentioned above.  Of those targeted objects,
we are able to secure redshifts for $>70$\%.  Follow-up observations
have shown that $\sim10$\% of the targets are objects at $z>1.5$
(C. Steidel, private communication).  We therefore have successful
redshifts for $\sim50$\% of all galaxies at $0.75<z<1.0$ in the
surveyed fields with apparent magnitude of $R<24.1$.  This sampling
rate is effectively scaled out when we dilute the mock galaxy catalogs
to compare to the DEEP2 data.  Further complications arise because the
sampling rate is  non-uniform due to the necessities of slitmask
design.  Spectra of objects are not allowed to overlap on the CCD,
such that objects that lie near each other in the direction on the sky
that maps to the wavelength direction on the CCD can not be
simultaneously observed; this results in modest ($\lesssim10$\%)
under-sampling in  regions with the highest density of targets on the
plane of the sky (each point on  the sky is covered by multiple DEEP2
slitmasks to ameliorate this effect).  We model this effect by
applying the actual DEEP2 maskmaking algorithm to the mock  galaxy
catalogs and then computing the VPF.  We find that the impact of the
target  selection algorithm on the VPF is negligible to within
$1\sigma$ (see Fig.~\ref{fig:mock_ccm}).   One would have expected
this, as the effect is most relevant on small scales and in overdense
regions.

There are also selection effects due to bright stars and other regions
that were not  observed.  These result in inhomogeneities on much
larger scales ($\gtrsim1$ Mpc) that could potentially be more relevant
for void studies.  To take account of these  effects for the DEEP2
sample, we generate an angular window function that has, for each
right ascension and declination, the completeness at that point
determined from the bad-pixel masks used in making our photometric
catalogs and the observed  redshift success for slitmasks covering
that point.  This window function therefore masks  unobservable
regions such as areas around bright stars (they are given a
completeness  of $0.0$).  We then convolve this window function with a
circular kernel  proportional to the path length through a sphere of
radius $R$ projected on the sky  (i.e., $K(\Delta\alpha,\Delta\delta)
\propto\sqrt{\rho^2-(\Delta\alpha)^2-(\Delta\delta)^2}$, where $\rho$
is the projected radius of the sphere in arcseconds, and
$\Delta\alpha$ and $\Delta\delta$ are the separations on the sky from
the center of the kernel in the right-ascension and declination
directions, respectively).

This convolution has a simple physical  consequence: we treat a test
sphere completely inside the survey volume at a  region of $50$\%
completeness as equivalent to a sphere only $50$\% in the survey
volume at a region of $100$\% completeness.  Our final product is a
window function  that contains, at each point, the projected
completeness averaged over a  sphere of a given radius, weighting by
the path length through the sphere.   We then throw down random
spheres only at points above a minimum convolved  completeness.  This
allows us to robustly avoid regions of bright stars, regions  of low
completeness (due, for example, to bad weather during observations)
and  the edges of the survey.   For DEEP2, we set the completeness
threshold at $55$\% so   that the allowable completeness range varies
over the survey by $\lesssim10$\%.   It is more important for our
purposes to be \emph{uniformly} complete than  highly complete, as an
overall dilution of the sample does not effect the reduced VPF.

\begin{figure}
\plotone{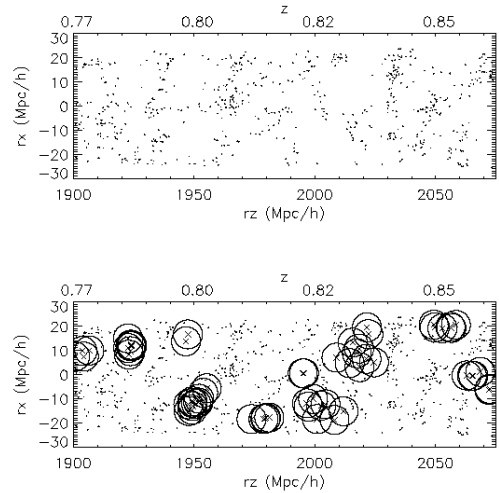}
\caption{Schematic demonstration of the statistical nature of the VPF.
The top panel shows  the comoving projected positions for galaxies in
a section of one DEEP2 field.  The suppressed  dimension is $\sim10$
$h^{-1}$ Mpc thick.  The bottom panel shows the empty spheres found
with a  search radius of $6$ $h^{-1}$ Mpc.  We show a very small
fraction of the actual number  of empty spheres detected at this
radius for clarity.}
\label{fig:deep2_show}
\end{figure}

We test the accuracy of this method by dividing the measured average
number of galaxies in a sphere of radius $R$, $\bar{N}$, by the sphere
volume for each radius.  We find that this inferred number density is
constant with sphere radius, providing good evidence that the larger
spheres do not tend to lay farther outside the survey geometry than
smaller spheres.  Finally, we have spot-checked by eye the locations
of the largest voids to ensure that they fall within the survey
geometry (see Fig.~\ref{fig:deep2_show}).

\subsection{Results at $z\sim1$}
We investigate the VPF and reduced VPF for DEEP2 galaxies in three
ways.  First we consider an ``overall'' sample consisting of all
galaxies with $-22<M_B-5\rm{log}(h)<-19$ and $0.75<z<1.0$.  Then we
compute the VPF as a function of galaxy color and of luminosity, using
volume-limited samples.  Errors for all DEEP2 measurements were
derived  from jackknife sampling using subvolumes of the three widely
separated DEEP2 fields.

\begin{figure}
\plotone{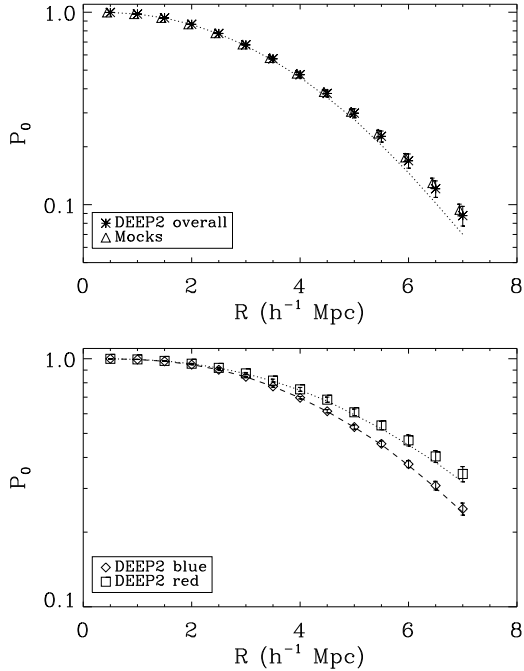}
\caption{\emph{Top}: $P_0$ for the overall DEEP2 sample compared to
mock galaxy catalogs.  The mock catalogs have an identical geometry as
the DEEP2 survey volume and have been passed through the  DEEP2 target
selection algorithm.  Both samples have the same number density
($n=0.006$  $h^3$ Mpc$^{-3}$).  The good agreement between the mock
catalogs and the data, although  on the surface encouraging, is due
entirely to the fact that the mock catalogs were constructed  to match
the DEEP2 two-point correlation function (see text for details).
\emph{Bottom}: DEEP2 $P_0$ separately for red and blue galaxies, both
diluted to  $n=0.002$ $h^3$ Mpc$^{-3}$.  The dilution removes any
number density effect on the VPF,  implying that we are only seeing
effects of clustering on the VPFs shown here.   Lines are predictions
of the negative binomial model.}
\label{fig:deep2_overall+colors}
\end{figure}

\begin{figure}
\plotone{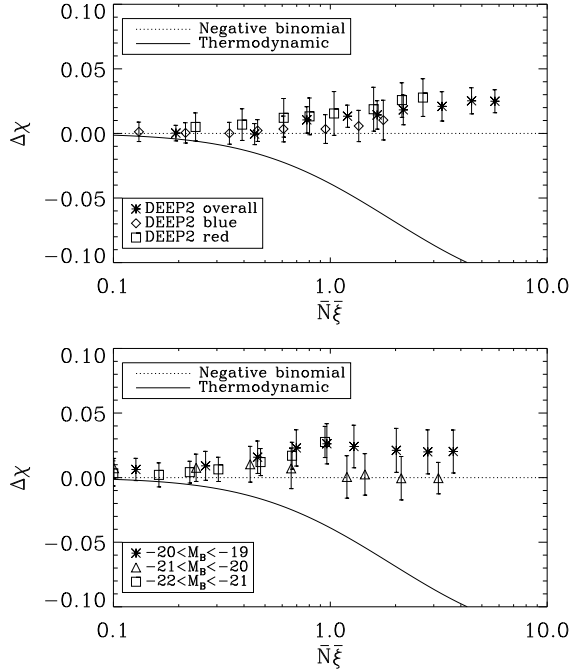}
\caption{Reduced VPF for DEEP2 galaxies, plotted as differences
between the negative binomial  model and the data.  \emph{Top}:
Differences for the overall, red, and blue samples.  The data  are
consistent with the model.  \emph{Bottom}: Differences for the DEEP2
sample in  differential magnitude bins.  Again there is good agreement
between the data and the  model.  The deviations on the largest scales
are likely insignificant due to the large covariance between bins,
which results in an underestimate of the true error.}
\label{fig:deep2_nb}
\end{figure}

We find excellent agreement between the overall DEEP2 sample VPF and
the mock galaxy catalog VPF (Fig.~\ref{fig:deep2_overall+colors}, top
panel).   The mock catalogs were randomly diluted to have the same
number density ($n=0.006$ $h^{3}$Mpc$^{-3}$) as the DEEP2 overall
sample.   The dashed line is a prediction of the negative binomial
model.   We show the reduced VPF, plotted as differences from the
negative binomial  prediction, for the overall sample in the top panel
of Fig.~\ref{fig:deep2_nb},  top panel.  The good agreement between
the data and the model implies that the VPF  can be described entirely
by $\bar{N}$ and $\bar{\xi}$.  The largest  deviations on large scales
seen here and below are of low statistical significance;   the data
are highly covariant from bin to bin, and hence our errors are
underestimates of the true error.  The excellent agreement between the
overall sample and mock  galaxy VPF reflects the fact that the mock
galaxy catalogs  were constructed to match the observed  $\bar{\xi}$
in the DEEP2 data.  Stated differently, since these samples have the
same number density (by construction), the fact that they have similar
VPFs implies  that they will have similar $\bar{\xi}$ values, and vice
versa.

We next divide the DEEP2 sample into galaxies with $(U-B)_0<1$ or
$>1$, roughly matching the  observed saddle point in the color
bi-modality.  We find that samples of  red galaxies have more and/or
larger voids than samples of blue galaxies
(Fig.~\ref{fig:deep2_overall+colors}, bottom panel; both samples were
randomly diluted to $n=0.002$ $h^{3}$Mpc$^{-3}$).   This dilution is
critical; without it, the differences between the red and blue galaxy
VPF   would be dominated by differences in the observed number density
of red versus blue galaxies.  It would be very difficult  to separate
the effects of number density from clustering strength  on the number
and size of voids if we simply compared the VPFs of undiluted samples.

As with the overall sample, we find that the reduced VPF of both the
blue  and red galaxy populations follows the negative binomial model
within the errors (Fig.~\ref{fig:deep2_nb}, top panel).  Again, this
implies that the differences between the VPFs for blue and red
galaxies (Fig.~\ref{fig:deep2_overall+colors}, bottom panel) are
solely due to differences in  their two-point correlation functions.
The similarity in reduced VPFs is somewhat  surprising.  We know that
blue and red galaxies are biased differently relative to the dark
matter \citep[e.g.,][and references therein]{Zehavi02,Coil04a,Coil05a}, and
that the  $S_p$ values are dependent on the bias.   We might have
expected the reduced VPFs,  which according to Eqn.~\ref{eqn:chi}
should depend on the $S_p$ values,  to be different for these two
populations.

The halo model affords a more direct interpretation of these results.
We have seen in $\S$~\ref{s:constrain} that the reduced VPF is
insensitive to  parameters of the HOD.  From studies of galaxy
correlations at $z\sim0$, it is becoming apparent  that different
populations of galaxies can have very different HODs.  Hence the
result  that red and blue galaxies have similar reduced VPFs is
consistent with  our tests using mock catalogs; very different HODs
will still produce the same  reduced VPF.

\begin{figure}[t!]
\plotone{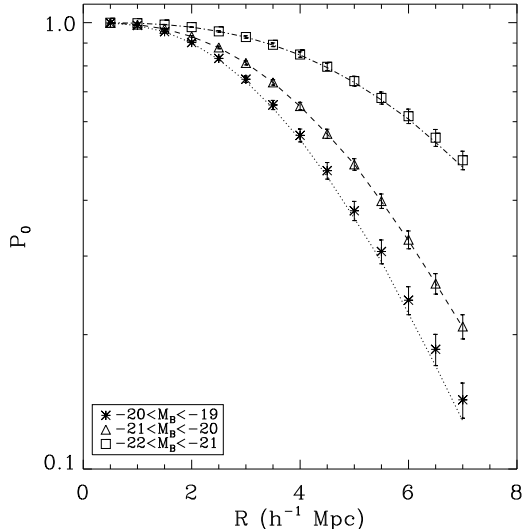}
\caption{Plot of $P_0$ for DEEP2 galaxies as a function of luminosity.
The competing effects  due to the number density and clustering
strengths drive the differences  between subsamples.  Given that the
reduced VPF for these subsamples can be  characterized by the negative
binomial model (see Fig.~\ref{fig:deep2_nb}), we  predict $P_0$ for
each sample using only the number density and volume-averaged
two-point correlation function measured within spheres of the given
radius (lines).   The excellent agreement seen here between the data
and the predictions  simply reflects the agreement found in
Fig.~\ref{fig:deep2_nb} (\emph{bottom}).}
\label{fig:deep2_mag}
\end{figure}

We also investigate the DEEP2 VPF dividing the sample into three
luminosity bins: $-20<M_B<-19$,  $-21<M_B<-20$, and $-22<M_B<-21$.  We
take a slightly different approach for this analysis.  Since the
number densities of the three luminosity subsamples  vary by roughly a
factor of four, the dilution required for  a direct comparison of VPFs
can discard a great deal of information and  unnecessarily increases
Poisson errors.  For this  reason, we begin by analyzing the reduced
VPF for each luminosity sample  (Fig.~\ref{fig:deep2_nb}, bottom
panel), again finding  good agreement with the negative binomial
model.  Using this model, we can  then use Eqn.~\ref{eqn:nbvpf} to
predict the VPF at any radius based solely on our knowledge of
$\bar{N}$ and $\bar{\xi}$ at that radius.  In Fig.~\ref{fig:deep2_mag}
we plot the VPF for  each of the three magnitude  bins (without
diluting the number density) and also show the predictions  based on
the negative binomial model; again, the agreement is within the
errors.   Knowledge of the form of the reduced VPF allows us to
separate the effects of number density from clustering on the VPF.

We note that the three magnitude bins cover slightly different
redshift intervals.  While the  faintest bin
($-20<M_B-5\rm{log}(h)<-19$) spans $0.7<z<0.85$, the brightest  bin
($-22<M_B-5\rm{log}(h)<-21$) spans $0.83<z<1.05$.  Each redshift
interval is chosen such  that the sample being considered should have
a roughly constant number density as a function of redshift.  This
ensures that there is no artificial redshift dependence in the void
distribution;  i.e., the samples are volume limited.  The less
luminous galaxies have a constant number density only over a limited
redshift range  because they have $R>24.1$ at higher redshifts.

The reduced VPF of all populations of galaxies at $z\sim1$ is well fit
by the negative binomial model.  Hence the VPF, a statistic that in
principle  relies on the infinite hierarchy of correlation functions,
can in fact be accurately described  solely by the number density and
the volume-averaged two-point correlation  function of the sample.  We
conclude that, at $z\sim1$,  the VPF provides no constraints on either
the halo model or on cosmological  parameters that cannot be gleaned
from studies of correlation statistics.

\section{Void Statistics from SDSS}\label{s:sdss}

\subsection{The Data}\label{s:sdss_data}
The Sloan Digital Sky Survey \citep[SDSS;][]{York00,Abazajian04} is an
 extensive photometric and spectroscopic survey  of the local
 universe.  Imaging data exist over $10^4$ deg$^2$ in five bandpasses,
 $u$,$g$,$r$,$i$, and $z$.  We use color conversions provided by
 M. Blanton (private communication) to derive $B$-band magnitudes in
 the AB system,  with typical 1-$\sigma$ errors in the conversion of
 0.2 magnitudes.   Approximately $10^6$ objects are being targeted for
 follow-up  spectroscopy as part of the SDSS; most spectroscopic
 targets are brighter than $r=17.77$ \citep{Strauss02}.  Automated
 software performs all the necessary  data reduction, including the
 assignment of redshifts.  Redshift errors are $\sim30$ kms$^{-1}$,
 similar to DEEP2 .  The spectrograph tiling algorithm ensures  nearly
 complete sampling \citep{Blanton03a}, yet the survey is not $100$\%
 complete due to several effects: (1)  fiber collisions that do not
 allow objects separated by $<1'$ to be simultaneously targeted,
 affecting $\sim6$\% of targetable objects, (2) a small fraction
 ($<1$\%) of  targeted galaxies fail to yield a reliable redshift, and
 (3) bright Galactic stars  block small regions of the sky.  Unlike
 the DEEP2 survey, (3) is not as  important for SDSS because a bright
 star will block out a much smaller comoving  volume at $z\sim0$
 compared to $z\sim1$.  The first of these effects is only important
 on  small scales, and is likely negligible for void statistics since
 an undersampled, intrinsically  high density region will not be
 counted as a void.

For this analysis we make use of the hybrid NYU Value Added Galaxy
Catalog (VAGC) \citep{Blanton05}.  This catalog combines the SDSS Data
Release 2 with a multitude of other publicly available  catalogs (2dF,
2MASS, IRAS PSCz, FIRST, and RC3), and includes a variety of derived
parameters including  $K$-corrections \citep{Blanton03b} and
structural parameters.

From the VAGC we have selected the two largest contiguous regions of
the SDSS.  We will call $SDSS1$ the region at $\alpha=190^o$,
$\delta=50^o$, which includes $\sim103,000$ galaxies with
spectroscopic $z<0.2$ and $SDSS2$ the region centered  on
$\alpha=190^o$, $\delta=1^o$ with $\sim87,000$ galaxies at $z<0.2$.
We divide red and blue galaxies in the SDSS data at the valley visible
in the rest-frame $g-r$ color distribution at $(g-r)=0.7$.

In addition to the `main' SDSS sample there is a secondary targeting
algorithm  designed to identify large numbers of luminous red galaxies
(LRGs) at moderate redshifts via  photometric color cuts.  These LRGs,
though low in spatial number density, cover an enormous volume, and
are hence  ideal objects for measuring very large scale clustering in
the universe \citep[see][for a more detailed description of the SDSS
LRG sample]{Zehavi05, Eisenstein05}.  The LRG sample we study here
contains $\sim10,000$ galaxies and spans the redshift range
$0.16<z<0.46$ and luminosity range $-23.2<M_g<-21.8$ with $(g-r)>0.7$
(note that this is the only case where we use SDSS absolute
magnitudes).

\subsubsection{Survey Geometry and Completeness}
Since the SDSS is a low-redshift survey, the angular size of a test
void sphere  of the same comoving radius varies enormously over the
redshift range $0.05<z<0.2$.  Paralleling our analysis of the DEEP2
sample, we account for geometry and  completeness effects in the
following way. We generate an angular window  function for the SDSS
with the aid of \emph{Mangle} \citep{Hamilton04} over  a dense grid in
right ascension and declination; the resulting resolution  is $0.15$
degrees in right ascension and declination.  Completeness values of
either 0 or 1 were assigned to each right point of this grid depending
on the  spectroscopic coverage and locations of bright stars (as SDSS
spectroscopy is  highly uniform in depth).  We then convolve this
window function with a kernel that represents the depth through the
test void projected on the sky (see  $\S$~\ref{sec:deep_compl} for
details) and only place random spheres in regions  above a minimum
convolved completeness.  This allows us to avoid placing test  spheres
in poorly-sampled regions.  We set this threshold at $85$\%, so that
the completeness within spheres placed down varies over the survey by
$\lesssim10$\%.

The net result of the strong scaling of angular size with redshift is
that  larger-volume test spheres cannot be placed in  the lowest
redshift bins because they would span a region of the sky comparable
to the  entire survey.  Hence at larger void radii we are restricted
to higher redshifts.   This does not cause a detrimental bias,
however, because we only consider galaxies over a redshift range such
that their number density is approximately constant,  i.e.,
volume-limited samples.  Hence a bias towards slightly higher
redshifts  ($z\sim0.1$) for larger voids will not skew our statistics,
assuming that there is not strong  void statistic evolution from
$z\sim0$ to $z\sim0.1$.  Indeed no significant evolution has been
detected in the VPF out to $z\sim0.3$ \citep{Hoyle04}.

There is one final effect in the SDSS data that, if not properly
treated, could significantly bias our results.  There is a massive
structure in $SDSS2$ at $z\sim0.08$ dubbed the ``Sloan Great Wall''
\citep{Gott05},  which will strongly affect any clustering
measurements.  This structure  has been removed in the correlation
function studies of \citet{Zehavi02},  and we do the same here by
defining our samples to avoid $0.075<z<0.085$  in the $SDSS2$ sample.
The structure extends across the entire angular  extent of $SDSS2$; if
included it would cause gross underestimates of the  cosmic error in
any large-scale structure measurement.  Based on many  realizations of
large cosmological simulations with Gaussian initial  conditions, it
is expected that a structure of this size occurs in a  volume the size
of the SDSS approximately $10$\% of the time \citep{Tegmark04}.   With
the full SDSS dataset, it should be possible to accurately account for
this enormous structure in the error budget without relying on
simulations.

In order to accurately measure quantities via our counts-in-cells
approach,  it is important that the number of randomly placed test
spheres exhausts the  number of independent volumes in the survey.
The second data release of  the SDSS over the interval $0.8<z<0.15$
spans a volume of $\sim 10^7$  $h^{-3}$ Mpc$^3$.  This corresponds to
$\sim 10^6$ independent volumes  for $1$ $h^{-1}$ Mpc spheres and
$\sim 10^3$ for $15$ $h^{-1}$ Mpc spheres, a  factor of $10$ more than
in the DEEP2 survey.  The number of spheres we use  always exceeds the
number of independent volumes.
 
\subsection{Results at $z\sim0$}
We now present void statistics in the local universe using the SDSS
dataset.   As in our analysis of DEEP2 data, we investigate void
statistics  for three sets of SDSS samples: an ``overall'' sample with
$-22<M_B-5\rm{log}(h)<-19$  and $0.09<z<0.14$, two subsamples split
according to $g-r$ color (at $(g-r)=0.7$),  and three subsamples
differing in luminosity.  In this section, as before, we plot reduced
VPFs as differences from the negative  binomial model in order to
clearly show small deviations.  As for the DEEP2 luminosity
subsamples,  All comparisons made in this section are made without
requiring that samples have similar number densities, and by using the
negative binomial model to predict the VPF (via Eqn.~\ref{eqn:nbvpf}).
Errors  for all quantities were derived from jackknife subsamples.

\begin{figure}[t!]
\plotone{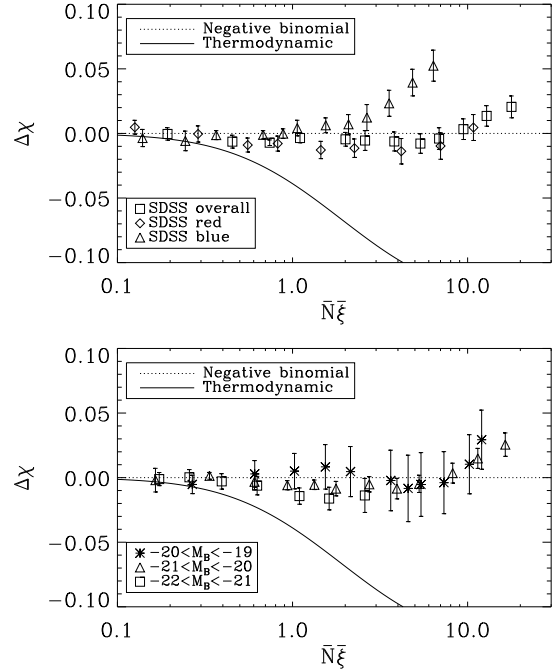}
\caption{Reduced VPF for SDSS galaxies, plotted as differences from
the negative  binomial  model.  \emph{Top}: The overall, blue, and red
galaxy samples agree  well with the negative binomial model.  The
largest discrepancy is for blue galaxies, although due to the large
covariance between bins, these errors are underestimates of the true
error.  \emph{Bottom}:  Reduced VPF  for SDSS galaxies as a function
of luminosity.  All luminosity  classes agree well with the model,
with the largest deviations seen in the  intermediate luminosity
sample. }
\label{fig:sdss_rvpf}
\end{figure}

\begin{figure}[t!]
\plotone{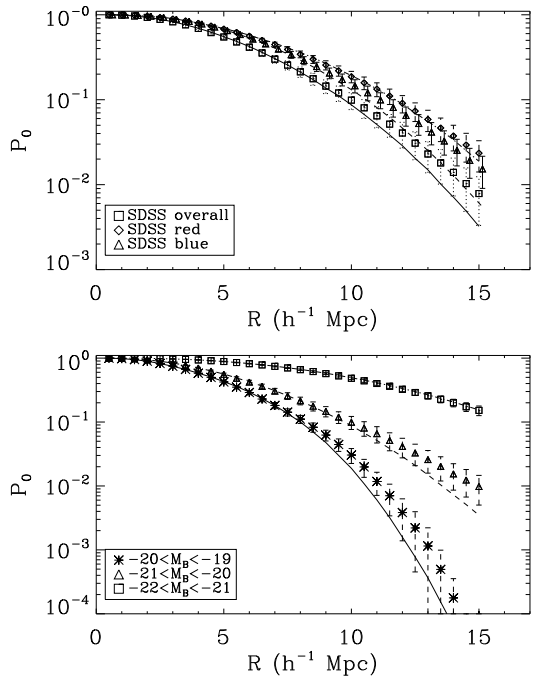}
\caption{Plot of $P_0$ measured for SDSS galaxies, overall and as a
function of color (\emph{top}) and  luminosity (\emph{bottom}).  The
data are shown as points,  while lines are predictions from the
negative binomial model for each subsample.   The predictions are
determined from measurements of $\bar{N}$ and $\bar{\xi}$ for  spheres
as a function of radius.   The agreement implies that $\bar{N}$ and
$\bar{\xi}$ alone determine $P_0$.  }
\label{fig:sdss_vpf}
\end{figure}

The reduced VPFs for the overall, red, and blue samples
(Fig.~\ref{fig:sdss_rvpf},  top panel) are all well described by the
negative binomial model.  As in the previous sections,  we can use
this measured agreement to predict the VPF at any radius based solely
on $\bar{N}$  and $\bar{\xi}$ for that radius, (e.g. as estimated from
counts-in-cells), for each sample.    Fig.~\ref{fig:sdss_vpf} (top
panel) plots the VPF for these samples and compares them to the
predictions from the negative binomial model (lines).  The trend that
the  red galaxy sample has more and/or larger voids than the blue
sample is due  to the fact that it is both more strongly clustered and
lower in number density than our sample of blue galaxies.  That the
measured data agree with the negative binomial model within errors
implies that, just as at $z\sim1$, the VPF cannot be  used to uniquely
constrain the halo model or cosmological parameters.  As with the
overall DEEP2 VPF, the blue SDSS data do not agree within $1\sigma$
errors  on the largest scales; this is likely insignificant, as the
data are highly covariant, and hence our errors are underestimates of
the true error.

Next, we compare void statistics for samples of SDSS galaxies as a
function of luminosity.  The reduced VPF for galaxies in all
luminosity bins we consider  ($-22<M_B-5\rm{log}(h)<-21$,
$-21<M_B-5\rm{log}(h)<-20$, and $-20<M_B-5\rm{log}(h)<-19$),  is in
good agreement with the negative binomial  model
(Fig.~\ref{fig:sdss_rvpf}, bottom panel).  We again  make this
agreement explicit by computing the VPFs for these samples and
comparing to  that predicted by assuming the negative binomial model
from the values of    $\bar{N}$ and $\bar{\xi}$
(Fig.~\ref{fig:sdss_vpf}, bottom panel).  Since these samples are
volume limited, each luminosity sample spans a slightly  different
redshift range, with the brightest spanning the largest range
($0.11<z<0.20$) (since bright galaxies can be detected at greatest
distances) and the faintest spanning  the smallest range
($0.05<z<0.07$).

\begin{figure}
\plotone{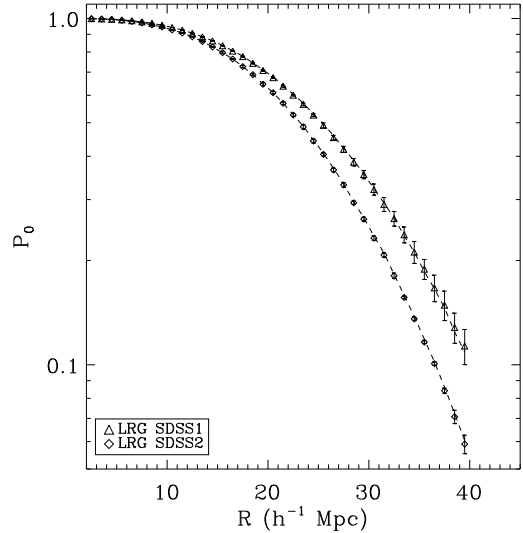}
\caption{Measured $P_0$ for the SDSS LRG population over the redshift
range $0.16<z<0.46$.  Unlike for the other SDSS data samples, here  we
are able to  measure $P_0$ to much larger radii and therefore show
results separately for  each of the two SDSS regions on the sky,
denoted $SDSS1$ and $SDSS2$  (see $\S$\ref{s:sdss_data}). The
difference between the  two regions reflects the cosmic variance of
their volumes.  As before, we show jackknife error estimates.  The
negative binomial predictions (lines)  match the measured $P_0$ values
well out to void radii of $40$ $h^{-1}$ Mpc.}
\label{fig:sdss_lrg}
\end{figure}

Finally, we compute the VPF for the LRG population, defined  to be
bright ($-23.2<M_g<-21.8$) and red ($(g-r)>0.7$), spanning the
redshift range $0.16<z<0.46$.  Their low number density
($n\sim10^{-5}$ $h^{3}$ Mpc$^{-3}$) and large redshift range allow for
a measurement of the VPF out to unprecedented  scales (R$=40$ $h^{-1}$
Mpc).  Fig.~\ref{fig:sdss_lrg} shows that the LRG population  is well
described by the negative binomial model out to the largest void radii
tested.  Unlike our analysis of other SDSS data, we plot the VPF
separately for the two  regions used in this study (denoted $SDSS1$
and $SDSS2$).  The difference between the  two regions reflects the
impact of cosmic variance on void statistics.

The reduced VPFs of all galaxy populations explored at $z\sim0$ are in
good agreement with  the negative binomial model.  It is quite
surprising that this agreement, which provides a simple  mapping from
$\bar{N}$ and $\bar{\xi}$ to $P_0$, is equally valid from $z\sim1$ to
$z\sim0$.   What this tells us is that the evolution in the VPF for a
given population from $z\sim1$ to $z\sim0$ must proceed in lockstep
with  evolution in the number density and two-point correlation
function of that population.  Finally, it is encouraging that these
conclusions at $z\sim0$ agree with \citet{Croton04}, who measure  void
statistics for the low-redshift 2dF survey, and find good agreement
between  the galaxy reduced VPF and the negative binomial model for a
range of galaxy luminosities.

\section{Discussion}\label{s:disc}

We have presented measurements of the void probability function (VPF)
and reduced VPF for galaxies as a function of color and luminosity at
$z\sim1$ using DEEP2 data and at $z\sim0$ with SDSS data.  We find
that all samples are well described by the negative binomial model.
This agreement implies that the VPF for a given sample is determined
entirely by the sample number density and volume-averaged two-point
correlation function, $\bar{\xi}$.  In particular, evolution of the
VPF for a population of galaxies from $z\sim1$ to $z\sim0$ is governed
by the evolution in $\bar{\xi}$ and $\bar{N}$ for that population.  We
have furthermore shown, using mock catalogs, that this simple relation
between the  VPF and $\bar{\xi}$ holds for a wide range of halo models
in redshift space, but breaks down in real space.  We now discuss the
relevance and implications of these findings.

The VPF we measure at $z\sim1$ is in good agreement with the VPF
measured in mock galaxy catalogs created from $\Lambda$CDM
simulations.  Although this is encouraging, our result that the
two-point  correlation function and number density determine the VPF
implies that this agreement found between data and simulations is
simply a consequence of the fact that the simulations were constructed
to match the number density and two-point correlation function  of
DEEP2 galaxies.  Stated differently, the VPF currently cannot provide
new  constraints either on cosmological parameters or on the method in
which  galaxies are placed into dark matter N-body simulations.

Previous void studies have interpreted the VPF according to its
mathematical expansion as an infinite sum of higher-order correlation
functions, using the hierarchical \emph{Ansatz} to relate the
higher-order functions to lower-order functions.   These studies then
attempt to use the reduced VPF to test the validity of this
\emph{Ansatz}.  Strictly speaking, however, finding that the VPF for
various populations of galaxies can be characterized entirely by
$\bar{N}$ and $\bar{\xi}$ does {\it not} provide evidence for the
validity of this \emph{Ansatz}.  In fact, the interpretation of the
VPF in terms of higher-order correlation functions implies that the
reduced VPF depends only on the scaling coefficients
($S_p\equiv\xi_p/\xi^{p-1}$), which suggests that populations with
different biases (and therefore different scalings  between the
three-point and two-point functions) should actually not have the same
reduced VPF.  We find just the opposite in the data, namely that
populations with different biases have the same reduced VPF.  The
interpretation of the VPF as a sum of correlation functions affords
little insight into these results.  Despite this,  the interpretation
predicts that the reduced VPF of a sample should be independent of
that samples number density, and this is indeed observed.

The halo model provides a more tractable theoretical framework.   We
have shown that the reduced VPF of mock galaxies is quite insensitive
to particular parameters of the HOD when voids are measured in
redshift space.  Using measurements  of $\xi$ to constrain the halo
model, it appears that blue and red  galaxies have very different HODs
\citep{Zehavi05b}.  The samples  we investigate here hence likely have
a range of HODs.  Yet,  based on the result that different HODs
generate the same reduced  VPF, it is not surprising that different
samples have the same reduced  VPF.

Furthermore, an analysis of the dark matter {\it halo} reduced VPF has
led us to conclude that it is the redshift-space distribution of halos
themselves that is primarily responsible for the agreement between the
measured reduced VPF and the negative binomial model.  The reduced VPF
changes very little when one populates dark matter halos with
galaxies.  It is only surprising that the minimum halo mass,
$M_{min}$, has very little effect on the reduced VPF as well.  This
insensitivity to $M_{min}$ implies that the changes in $\bar{\xi}$
caused by changes in $M_{min}$ are enough to completely account for
the changes in the VPF.

We would like to stress that the strict dependence of the VPF on
$\bar{N}\bar{\xi}$ does not rely on any theoretical  interpretation,
including the hierarchical \emph{Ansatz}, and, in particular,  does
not depend on the reduced VPF following the negative binomial model.
Our result that all galaxy populations studied here have consistent
reduced VPFs immediately implies that the VPF can be entirely
described by  $\bar{N}\bar{\xi}$.  This, in turn, implies that  the
VPF is currently incapable of uniquely providing constraints on either
cosmological parameters or particular aspects of the halo model, as
any useful  information provided by the VPF is likewise provided by
two-point correlation  function analyses; the VPF is of Little use in
understanding the large scale  structure of the universe.

\acknowledgments  We would like to thank Andreas Berlind, Michael
Blanton, Darren Croton, Ofer Lahav,  Chung-Pei Ma, Roman Scoccimarro,
and Simon White for  useful discussions.  This project was supported
in part by the NSF grants AST00-71198 and  AST00-71048.  The DEIMOS
spectrograph was funded by a grant from CARA  (Keck Observatory), an
NSF Facilities and Infrastructure grant (AST92-2540),  the Center for
Particle Astrophysics, and gifts from Sun Microsystems and  the
Quantum Corporation.  C.C. acknowledges support from an NSF REU grant.
J.N. acknowledges support from NASA through Hubble  Fellowship grant
HST-HF-01165.01-A awarded by the Space Telescope Science  Institute,
which is operated by the Association of Universities for Research  in
Astronomy, Inc., for NASA, under contract NAS 5-26555.  Some of the
data  presented herein were obtained at the W.M. Keck Observatory,
which is operated  as a scientific partnership among the California
Institute of Technology,  the University of California, and the
National Aeronautics and Space  Administration. The Observatory was
made possible by the generous financial  support of the W.M. Keck
Foundation.  In addition, we wish to acknowledge  the significant
cultural role that the summit of Mauna Kea plays within  the
indigenous Hawaiian community; we are fortunate to have the
opportunity  to conduct observations from this mountain.

Funding for the creation and distribution of the SDSS Archive has been
provided  by the Alfred P. Sloan Foundation, the Participating
Institutions, the National  Aeronautics and Space Administration, the
National Science Foundation, the U.S.  Department of Energy, the
Japanese Monbukagakusho, and the Max Planck Society.  The SDSS Web
site is http://www.sdss.org/.

The SDSS is managed by the Astrophysical Research Consortium (ARC) for
the  Participating Institutions. The Participating Institutions are
the University  of Chicago, Fermilab, the Institute for Advanced
Study, the Japan Participation  Group, Johns Hopkins University, the
Korean Scientist Group, Los Alamos  National Laboratory, the
Max-Planck-Institute for Astronomy (MPIA), the  Max-Planck-Institute
for Astrophysics (MPA), New Mexico State University,  the University
of Pittsburgh, the University of Portsmouth, Princeton University, the
United States Naval Observatory, and the University of Washington.

%\bibliography{master_refs}

\end{document}